\documentclass{article}
\usepackage[dvips]{graphicx}
\usepackage{amssymb}
\usepackage{amsmath}
\usepackage{amscd}
\usepackage{mathrsfs}
\usepackage{subfigure}
\usepackage{latexsym}
\usepackage{makeidx}
\date{}
\author{Tina A.C. Maiolo\footnote{Dipartimento di Fisica dell'Universit\`a and Sezione INFN, 73100 Lecce, Italy; e--mail: tina.maiolo@le.infn.it}, Fabio Della Sala\footnote{NNL and CNR--INFM of Lecce, 73100 Lecce, Italy; e--mail: fabio.dellasala@unile.it}, Luigi Martina\footnote{Dipartimento di Fisica dell'Universit\`a and Sezione INFN, 73100 Lecce, Italy; e--mail: Martina@le.infn.it}, Giulio Soliani\footnote{Dipartimento di Fisica dell'Universit\`a and Sezione INFN, 73100 Lecce, Italy; e--mail: Soliani@le.infn.it}}
\title{\textbf{Entanglement of electrons in interacting molecules}}

\begin{document}
\maketitle
\begin{abstract}
\noindent
Quantum entanglement is a concept commonly used with reference to the existence of certain correlations in quantum systems that have no classical interpretation. It is a useful resource  to enhance the mutual information of memory channels or to accelerate some quantum processes as, for example, the factorization in Shor's Algorithm. Moreover, entanglement is a physical observable directly measured by the von Neumann entropy of the system. We have used this concept in order to give a physical meaning to the electron correlation energy in systems of interacting electrons. The electronic correlation is not directly observable, since it is defined as the difference between the exact ground state energy of the many--electrons Schr\"odinger equation and the Hartree--Fock energy.
We have calculated the correlation energy and compared with the entanglement, as functions of the nucleus--nucleus separation using, for the hydrogen molecule, the Configuration Interaction method.  
 Then, in the same spirit, we have analyzed a dimer of ethylene, which represents the simplest organic conjugate system, changing the relative orientation and distance of the molecules, in order to obtain the configuration corresponding to maximum entanglement.

\vspace{.5cm}
\noindent 
\textbf{Key Words}. entanglement; electron correlation energy; interacting molecules. 
\end{abstract}

\section{Introduction}
Much attention has been recently devoted to the notion of quantum entanglement for its potential applications in quantum information theory \cite{nie00}. Indeed, quantum entanglement holds a fundamental role in quantum teleportation \cite{ben93}, superdense coding \cite{ben92}, quantum key distribution \cite{eke91}, quantum cryptography \cite{fuc97}, and has been proved to be necessary for an exponential speed--up of quantum computation over classical computation.

 For these reasons, a lot of efforts have been devoted to characterize entanglement and experimental measurements have demonstrated that it can affect the macroscopic properties of solids \cite{gho03}. For example, Chen  \emph{et al.} analyzed in \cite{che06} entanglement of the ground states in XXZ and dimerized Heisenberg spin chains as well as in two--leg ladder by using the spin--spin concurrence and the entanglement entropy between a selected sublattice of spins and the rest of the system. He \emph{et al.} showed in \cite{lix05} the energy spectrum, pair--correlation functions and the degree of entanglement of two--electron states in self--assembled InAs/GaAs quantum dot molecules via all--bound--state configuration interaction calculation and compared these quantities in different levels of approximations. Buscemi \emph{et al} performed in \cite{bus06} the quantitative evaluation of the entanglement dynamics in scattering events between two indistinguishable electrons interacting via Coulomb potential in 1D and 2D semiconductor nanostructures. They followed the idea introduced by Schliemann \cite{sch01} and extended by other outhors \cite{ghi04}, \cite{hua05}, since it allows to use a density matrix in the spatial coordinates and to compute the von Neumann entropy. Such an approach is based on an analogous of the Schmidt decompostion for state vectors of two fermionic particles: through an unitary transformation the antisymmetric wave function is expressed into a basis of Slater determinants with a minimum number of non--vanishing terms. This number, known as Slater rank, is a criterion to identify whether a system is entangled or not, which involves the evaluation of the von Neumann entropy of one particle reduced density matrix. In order to exploit entanglement in quantum computation also more complex molecules are studied. For example, the researchers of Hahn--Meitner--Institut are working on the realization of molecular quantum computers whose quantum bits could be formed by fullerenes with a single nitrogen or phosphorus atom trapped inside. 

For this reason, it is important to investigate the properties of some molecules and the interaction among them. A study of this kind is carried out in this paper. Here, the starting point is the so--called Collin's conjecture \cite{nat93}, that is, the correlation energy in molecular systems is proportional to their entropy. This conjecture was confirmed by numerical evidence by Ramirez \emph{et al.} \cite{ram97} and taken up by Huang and Kais in \cite{hua05} who interpreted the degree of entanglement as an evaluation of correlation energy. Entanglement is in fact  a physical observable directly measured by the von Neumann entropy of the system. This concept is used in order to give a physical meaning to the electron correlation energy, which is not directly observable since it is defined as the difference between the total energy of a given molecular system, with respect to the one obtained with the Hartree--Fock approximation method. The Hartree--Fock method, in fact,  is typically used to solve Schr\"odinger equation for a multi--electron atom or molecule described in the fixed--nuclei approximation (Born--Oppenheimer approximation) by the electronic molecular Hamiltonian and the calculation of the error due to this approximation is a major problem in many--body theory and a vast amount of work has been done on this subject \cite{sza89}.

In this work, we make a measurement of electron entanglement in two different examples of bipartite systems, as $H_{2}$ molecule and the dimer of ethylene, where each hydrogen atom or each ethylene molecule, respectively, can be considered a qubit.
In this article, firstly, we decide the method we use to make a measurement of entanglement comparing the von Neumann entropy of the reduced density matrix $S(\rho_{1})$ and the von Neumann entropy of the density matrix $S(\rho)$ for the $H_{2}$ molecule. After, fixed that the most convenient method to adopt is the second, $(S(\rho))$, we make a measurement of entanglement in a many--body system represented by a dimer of ethylene and we compare it with its electron correlation energy. Finally, using the Klein's inequality for a bipartite system, we define the interaction electron entanglement in order to study the degree of entanglement between the two molecules of the dimer.

The article is organized as follows: in \textbf{Sec. 2} we  briefly resume what entanglement is \cite{sch35} (equivalently, when a state is separable or entangled \cite{doh02}) and why we choose the von Neumann entropy to make an estimate of the degree of entanglement. In \textbf{Sec. 3} we first analyze a dimer of ethylene considering it as a multielectrons system, calculate the correlation energy, and then compare it with the entanglement, as a function of the distance between the two molecules.
In \textbf{Sec. 4} we introduce a new quantity, the \emph{interaction electron entanglement}, which is defined as $S_{int}=S[2C_{2}H_{4}]-2S[C_{2}H_{4}]$, in order to obtain the entanglement between the two molecules of the dimer. Then, we change the relative orientation and the distance between the molecules, in order to obtain the configuration corresponding to maximum entanglement. In this way, the system can be considered bipartite and each molecule can be seen as a qubit for an application to quantum computing.
In \textbf{Sec. 5} we explain some computational details describing the package we use and how we prepare the input for the compound systems that we analyzed. Finally, in \textbf{Sec. 6} some comments and results are discussed.
\section{A measurement of entanglement}
In quantum mechanics one generally distinguishes between \emph{pure states} and \emph{mixed states} depending on the information one has on the state in which a physical system actually is \cite{des76}.

Pure states are usually introduced whenever one wants to describe a pysical system that is in a given state with certainty, and are represented by unit vectors $|\psi\rangle$ of the Hilbert space associated with the system. Equivalently a pure state can be represented by a density operator $\rho = | \psi \rangle \langle \psi|$ which is such that $\mathcal Tr \rho = \mathcal Tr \rho^2 = 1 $.

Whenever one doesn't  know exactly in which state the system is, but there is only a probability $p_{i}$ for the system to be in the pure state represented by $|\psi _{i}\rangle$, then one tipically says that the system is in a mixed state and represents it by the density operator 
$\rho = \sum _{i}p_{i}|\psi_{i} \rangle \langle \psi_{i}|= \sum_{i}p_{i}\rho_{i}$,
where $p_{i}\geqslant 0$, $\sum_{i}p_{i}=1$, and $\rho_{i}$ is the density operator associated with the pure state $|\psi_{i}\rangle$. Contrary to a pure state, a mixed state is such that 
$\mathcal Tr \rho  = 1\neq \mathcal Tr \rho^2$.

The most important consequences of the mathematical structure of the state space reveal in the study of compound systems. In fact, let us consider a physical system $T$ made up by the two subsystems $A$ and $B$ whose states are represented by the elements of the Hilbert spaces $\mathscr H_{A}$ and $\mathscr H_{B}$, respectively. Then, the system $T$ is described by $\mathscr H_{T} = \mathscr H_{A} \otimes \mathscr H_{B}$ and a pure state of $T$ can be represented by the unit vector $|\psi\rangle \in \mathscr H_{T}$. Let $\{|\phi_{i}(A)\rangle\}_{i\in \mathbb{N}}$ and $\{|\chi_{j}(B)\rangle\}_{j\in \mathbb{N}}$ be orthonormal bases on the Hilbert space $\mathscr H_{A}$ and   $\mathscr H_{B}$, respectively. A pure state of $\mathscr H_{T}=\mathscr H_{A}\otimes \mathscr H_{B} $ can be expressed in the form $|\psi\rangle =\sum_{i,j}c_{ij}|\phi_{i}(A)\rangle\otimes |\chi_{j}(B)\rangle$, where $\sum_{i,j}|c_{ij}|^{2}=1$.
Thus, if the state $|\psi\rangle \in \mathscr H_{T}$ can be written as a tensor product of an element of $\mathscr H_{A}$ and an element of $\mathscr H_{B}$, as in the following,
$|\psi\rangle =|\phi (A) \rangle \otimes |\chi (B) \rangle \equiv |\phi (A),\chi (B)\rangle$,
we say that $|\psi \rangle$ is a \emph{product} (or \emph{separable}, or \emph{factorizable}) pure state. A pure state of a pair of quantum systems is called entangled if it is unfactorizable.
According to a well known theorem due to von Neumann \cite{von32}, the vector $|\psi \rangle \in \mathscr H_{T}$ can be expressed in the so--called \emph{Schmidt decomposition}:
$|\psi\rangle=\sum_{i\geqslant 1}\sqrt{p_{i}}|\nu_{i}(A)\rangle|\lambda_{i}(B)\rangle$. Then, $|\psi\rangle$ is a product state if there is only one Schmidt coefficient different from zero (if $i=1$ in the sum). On the contrary, if there are more than one non--zero Schmidt coefficients (if $i>1$ in the sum) the pure state is entangled.

As a simple bipartite system, let us consider the $H_{2}$ molecule, whose Hilbert space can be represented by:
\begin{equation}
\mathscr H=\Big[\mathscr L^{m}(1) \otimes \mathscr S^{2}(1)\Big]\otimes \Big[\mathscr L^{m}(2) \otimes \mathscr S^{2}(2)\Big]=\mathscr C^{2m}(1)\otimes\mathscr C^{2m}(2),
\end{equation}
where $\mathscr L$ and $\mathscr S$ represent the orbital and the spin space, respectively; in brackets we denote one of the two electrons while the index represents the dimension of the space.
 
In the occupation number representation \cite{git02} $(n_{1}\uparrow, n_{1}\downarrow, n_{2}\uparrow, n_{2}\downarrow,\dots,n_{m}\uparrow, n_{m}\downarrow  )$ the subscripts denote the spatial orbital index and with this notation let us introduce an orthonormal basis for each space $\mathscr C^{m}$:
\begin{equation}
\left\{ \begin{array}{c}
|n_{1}\uparrow\rangle\\
|n_{1}\downarrow\rangle\\
|n_{2}\uparrow\rangle\\
|n_{2}\downarrow\rangle\\
\vdots\\
|n_{m}\uparrow\rangle\\
|n_{m}\downarrow\rangle
\end{array}\right\}_{(1)}\otimes\hspace{0.3cm}
\left\{ \begin{array}{c}
|n_{1}\uparrow\rangle\\
|n_{1}\downarrow\rangle\\
|n_{2}\uparrow\rangle\\
|n_{2}\downarrow\rangle\\
\vdots\\
|n_{m}\uparrow\rangle\\
|n_{m}\downarrow\rangle
\end{array}\right\}_{(2)}
\end{equation} 
A pure two-electron state $|\Psi\rangle$ can be written in this representation as 
\begin{equation}
|\Psi\rangle=\sum_{a=1}^{m}\sum_{b=1}^{m}\omega_{a,b}c_{a}^{\dag}c_{b}^{\dag}|0\rangle,
\end{equation}
where $|0\rangle$ is the vacuum state, the coefficients $\omega_{a,b}$ satisfy, accordingly to Pauli exclusion principle the following requests:
\begin{equation}
\left\{ \begin{array}{l}
\omega_{a,b}=-\omega_{b,a}\\
\omega_{i,i}=0,
\end{array}\right.
\end{equation}  
and $c^{\dag}$ and $c$ are the creation and annihilation operators of single--particle states, respectively, whose action on the vacuum state is
\begin{equation}
c_{a}^{\dag}c_{b}^{\dag}|0\rangle = |ab\rangle,\hspace{0.3cm} a,b\in \{ 1,2,3,4, \dots, 2m\}\hspace{1.0cm}
\begin{tabular}{llll}
$1\equiv |n_{1}\uparrow \rangle$ & $2\equiv |n_{1}\downarrow \rangle$ \\ 
$3\equiv |n_{2}\uparrow \rangle$ & $4\equiv |n_{2}\downarrow \rangle$ \\ 
\vdots & \vdots\\
$m-1\equiv |n_{m}\uparrow \rangle$ & $m\equiv |n_{m}\downarrow \rangle$\\
\end{tabular}
\end{equation}
The coefficients $\omega_{a,b}$ can be calculated by using the configuration interaction method \cite{pal82}, \cite{scu87}. In particular, for $H_{2}$ molecule we use configuration interaction single--double (CISD) method that is limited to single and double excitations:
\begin{equation}
|\Psi^{CISD}\rangle=c_{0}|\psi_{0}\rangle + \sum_{ar}c_{a}^{r}|\psi_{a}^{r}\rangle +\sum_{a<b,r<s}c_{a,b}^{r,s}|\psi_{a,b}^{r,s}\rangle,  
\end{equation}
where $|\psi_{0}\rangle$ is the ground state Hartree--Fock wave function, $c_{a}^{r}$ is the coefficient for single excitation from orbital $a$ to $r$, and $c_{a,b}^{r,s}$ is the double excitation from orbital $a$ and $b$ to $r$ and $s$. 
In the occupation number representation, the CISD wave function is given by
\begin{equation}
|\Psi\rangle= c_{0}|1_{1}1_{2}0_{3}0_{4}\dots\rangle +c_{1}^{3}|0_{1}1_{2}1_{3}0_{4}\dots\rangle + c_{2}^{4}|1_{1}0_{2}0_{3}1_{4}\dots\rangle + c_{1,2}^{3,4}|0_{1}0_{2}1_{3}1_{4}\dots\rangle+ \cdots  
\end{equation}
where each ``1'' represents the presence of the electron and the subscript represents the site where the electron is \cite{zan02}.

In order to realize whether the state of the compound system is entangled or not, we have to construct the corresponding reduced density matrix.
Indeed, the vector $|\psi\rangle$ represents a product state of the compound system if and only if the corresponding reduced density operators represent pure states (equivalently, $|\psi\rangle$ is an entangled state if and only if the corresponding reduced density operators formally represent mixed states).
Hence, if we want to measure the degree of entanglement we can calculate the degree of mixedness of the corresponding reduced density operator (or matrix) \cite{ben96}. This is a consequence of the fact that an entangled state gives rise to correlations whose information is lost if we only observe the subsystems, as it results from the fact that we effectively have a mixed state.

Of particular significance for describing randomness of a state and mixedness of density operators is the \emph{von Neumann entropy}, defined as
\begin{equation}
\label{entropy}
S(\rho)=-\mathcal Tr(\rho log_{2}\rho).
\end{equation}
It is important to note that $\rho$ is a non--negative trace class operator, while $S$ is not. In analogy with classical entropy, $S$ measures the amount of randomness in the state $\rho$.
More precisely, the entropy $S$ is zero if and only if the state is pure, and it is maximized when $S(\rho)=log_{2}d$,  where $d$ is the dimension of the Hilbert space. In other words, the more mixed is the reduced density operator, the more entangled is the original state  and  this result can be seen as a justification for the use of entropy as a measurement of quantum entanglement.
Thus, the entanglement of a pure state of a pair of quantum systems can be obtained as entropy of either member of the pair, consequently, in order to calculate the degree of electron entanglement in the $H_{2}$ molecule we have to calculate the von Neumann entropy of the density matrix reduced with respect to an electron. Starting from the density matrix $\rho=|\Psi^{\textsc{CISD}}\rangle\langle\Psi^{\textsc{CISD}}|$, we obtain the reduced density matrix by making the partial trace relative to all the occupation numbers except $n_{1} \uparrow$ (see Appendix A)  
\begin{displaymath}
\label{reduced}
\rho_{1}^{\textsc{CISD}}=Tr_
{
\begin{array}{c}
n_{i}\uparrow,n_{j}\downarrow \\
\begin{tiny}
\left\{ \begin{array}{c}
1< i \le m \\
1 \le j \le m
\end{array} \right.\end{tiny}
\end{array}
}
\rho^{\textsc{CISD}}=
\end{displaymath}
\begin{displaymath}
=\sum_{\begin{footnotesize}
\begin{array}{c}
n_{i}\uparrow=0,1\\ 
n_{j}\downarrow=0,1
\end{array}\end{footnotesize}
}
\langle n_{1}\downarrow,n_{2}\uparrow, n_{2}\downarrow,\dots ,n_{m}\uparrow, n_{m}\downarrow|\rho|n_{1}\downarrow,n_{2}\uparrow, n_{2}\downarrow,\dots ,n_{m}\uparrow, n_{m}\downarrow\rangle 
\end{displaymath}

\begin{equation}
\label{reddensH}
\rho_{1}^{CISD}=\left( \begin{array}{cc}
\sum_{i=1}^{m-1}|c_{1}^{2i+1}|^{2}+\sum_{i,j=1}^{m-1}|c_{1,2}^{2i+1,2j+2}|^{2} & 0\\
0 & |c_{0}|^{2}+\sum_{i=1}^{m-1}|c_{2}^{2i+2}|^{2}
\end{array}\right).
\end{equation}
The von Neumann entropy of the reduced density matrix $\rho^{CISD}$ represents the degree of entanglement:
\begin{equation}
S\Big(\rho_{1}^{CISD}\Big)=-Tr\Big(\rho_{1}^{CISD}log_{2}\rho_{1}^{CISD}\Big)=
\end{equation}
\begin{displaymath}
=-\Big(\sum_{i}^{m-1}|c_{1}^{2i+1}|^{2}+\sum_{i=1}^{m-1}|c_{1,2}^{2i+1,2i+2}|^{2}\Big)log_{2}\Big(\sum_{i}^{m-1}|c_{1}^{2i+1}|^{2}+\sum_{i=1}^{m-1}|c_{1,2}^{2i+1,2i+2}|^{2}\Big)+
\end{displaymath}
\begin{equation}
-\Big( |c_{0}|^{2}+\sum_{i=1}^{m-1}|c_{2}^{2i+2}|^{2}\Big)log_{2}\Big( |c_{0}|^{2}+\sum_{i=1}^{m-1}|c_{2}^{2i+2}|^{2}\Big),
\end{equation}
In order to make a generalization of this method to multipartite systems, we decide to measure the degree of entanglement as the von Neumann entropy of the density matrix corresponding to the compound system:
\begin{equation}
S\Big(\rho\Big)=-Tr\Big(\rho log_{2}\rho\Big)=-\frac{1}{2}\Big(\sum \alpha_{i}log_{2}\alpha_{i}+\sum\beta_{i}log_{2}\beta_{i}\Big),
\end{equation}
where $\alpha_{i}$ and $\beta_{i}$ are the so called \emph{Natural Spin Orbitals} (the eigenvalues of $\rho$).
We compare the behaviours of the entanglement obtained with the CISD coefficients method or with the Natural Spin Orbital method, calculated as a function of internuclear distance $R$. 

\begin{tabular}{cp{4.3cm}}           
\includegraphics[angle=-90, width=0.55\textwidth]{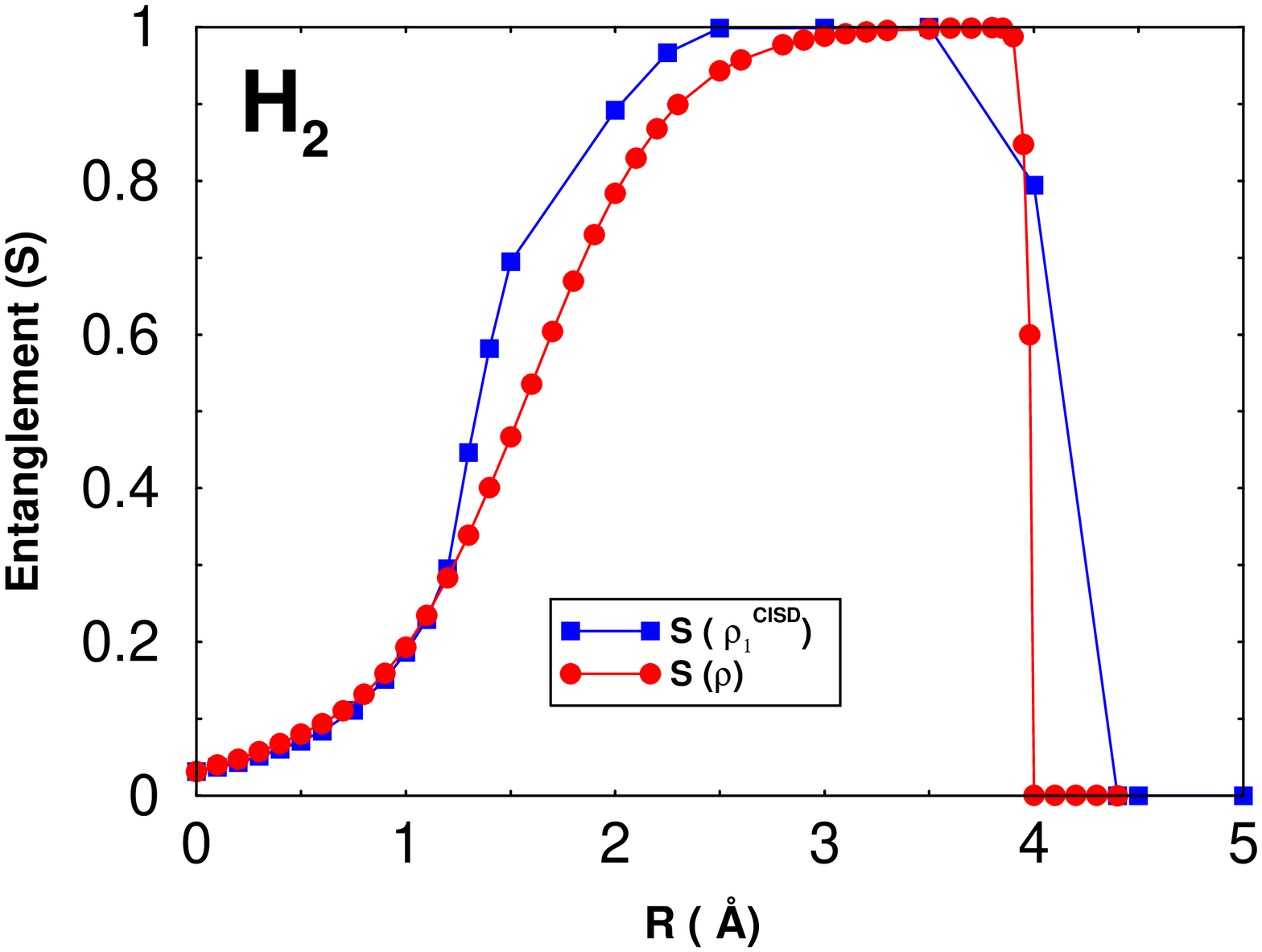}
&
{\vspace{0.3cm}\it{Figure 1: The figure shows the degree of entanglement in the Hydrogen molecule calculated as the von Neumann entropy of the reduced density matrix with respect to an electron $(S(\rho_{1}^{CISD}))$ and as the von Neumann entropy of the density matrix $(S(\rho))$.}}
\end{tabular}

In Figure 1 we show that the entanglement calculated as the von Neumann entropy of the reduced density matrix $S(\rho^{CISD}_{1})$ has a similar behaviour as the entanglement calculated as the von Neumann entropy of the density matrix $S(\rho)$ of the compound system. In particular, at small internuclear distances, the two curves representing the entropies overlap, while outside the bond region, they give different results though their behaviours are similar. Since we are interested to a qualitative course of entanglement and to the search of its maximum value, we neglect the above differences and choose to adopt the Natural Spin Orbital method in order to investigate more complicated systems. 
\section{Entanglement as a measurement of electron correlation} 
Our main aim in this Section is to show that entanglement can be interpreted as a measure of the electron correlation \cite{got05}, \cite{gro94}, \cite{ger97}. Entanglement, in fact, is a physical observable directly measurable by the von Neumann entropy of the system. On the contrary, the electron correlation energy is not directly observable, since it is defined as the difference between the exact total energy of a given molecular system, with respect to the one prescribed by the Hartree--Fock approximation method, that is
\begin{equation}
\label{correlation}
E_c=E_{exact}^{Sch}-E_{HF}.
\end{equation} 
 The correlation energy is the energy recovered by fully allowing the electrons to avoid each other and Hartree--Fock method improperly treats interelectron repulsions in an averaged way.
In other words, the exact wave function for a system of many interacting electrons is never a single determinant or a simple combination of a few determinants. The corresponding error is due to the correlations that are the analogue of the quantum entanglement in separated systems and are essential for quantum information processing in nonseparated systems.

Since in \cite{hua05} the authors discussed the formation of $H_{2}$ molecule, showing a qualitative agreement between the entanglement and the correlation energy as functions of nucleus-nucleus distance, let us generalize the argument proposed above to more complex molecules, as a dimer of ethylene.
Set the distance between the molecular planes, we calculate the electron correlation energy of the dimer by the procedure described in \textbf{Sec. 5}.
 For the same configuration, we calculate the degree of entanglement with the Natural Spin Orbitals method. We repeate the same procedure changing the distance between the planes or, once we have fixed this distance, we make a rotation as in picture and the results are in the first and in the second panel of Figure 2 respectively.

\begin{figure}
 \begin{minipage}[b]{0.5\textwidth}
   \centering
   \includegraphics[angle=-90, width=1.05\textwidth]{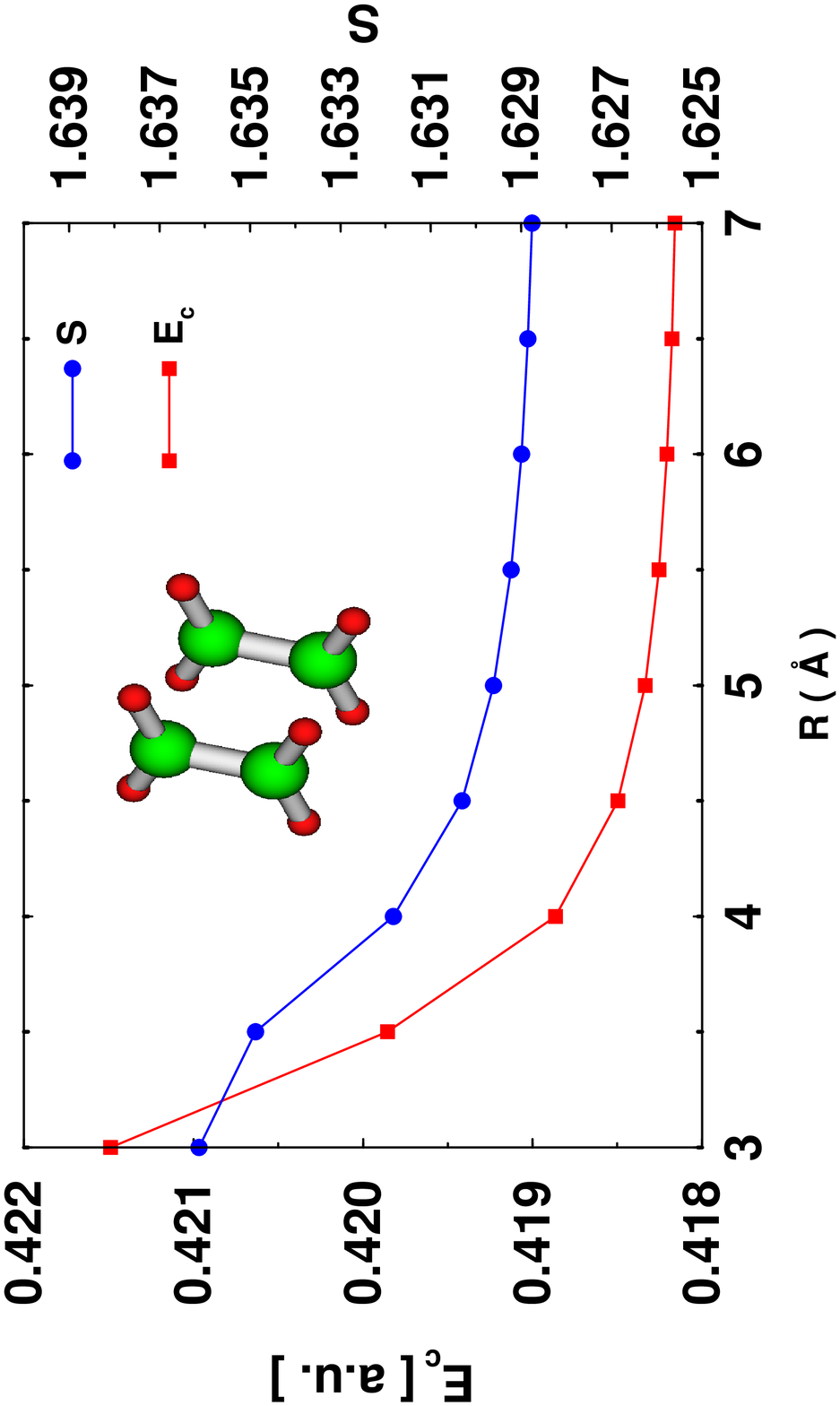}
 \end{minipage}
 \begin{minipage}[b]{0.5\textwidth}
   \centering
   \includegraphics[angle=-90, width=1.05\textwidth]{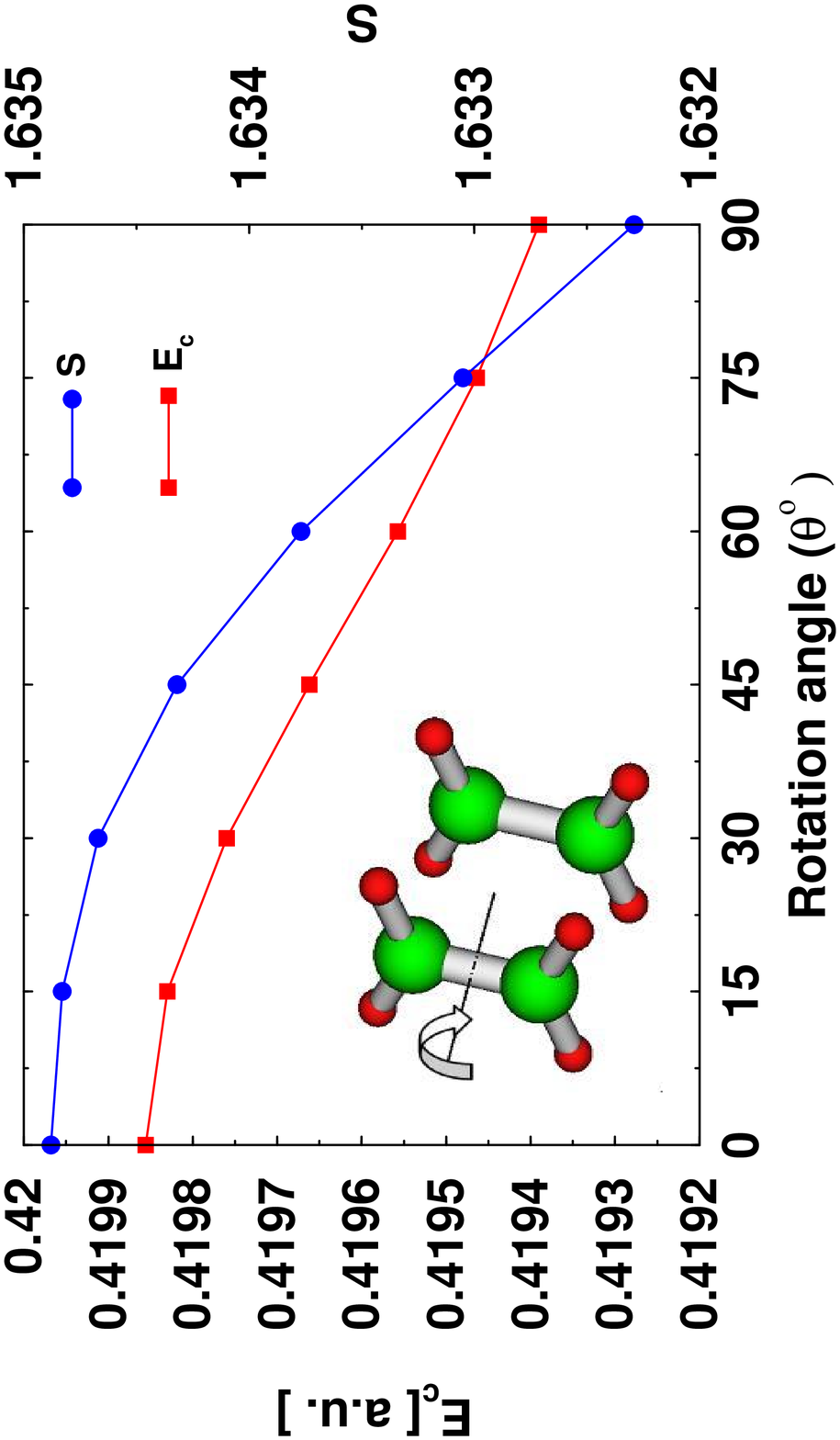}
 \end{minipage}
\it{Figure 2: A comparison between entanglement and correlation energy in the dimer of ethylene, calculated by changing the mutual distance of the molecules (left panel) with molecules in a face to face orientation and by changing the relative orientation (right panel, the molecules are parallel at R=3.5 \AA).}
\end{figure}

As we can see from the above picture, the entropy (S) and the correlation energy ($E_{c}$) have a comparable behaviour, although we use two different scales to represent these two quantities. In fact, when we plot them as functions of the plane distance, the two curves quickly decrease till $R=4.5 $ \AA, moreover for bigger distances, both of them decrease more slowly. When we set the distance at $R=3.5$ \AA $\,$ and change the mutual orientation of the molecules on their planes, we can see that both the entanglement and correlation energy decrease when the rotation angle grows up. The reason of this fact is that when a rotation of a molecule on its plane is made, the bigger is the rotation angle, the small becomes the superposition of the external orbitals of the two molecules. Consequently, the electron entanglement and the correlation energy decrease as it is shown in Figure 2.

The above reasonings suggest us to interpret the entanglement as an efficient instrument to evaluate the electronic correlation energy, not only for $H_{2}$ molecule \cite{hua05}, but also for some bigger systems.  

\section{Results: interacting molecules} 
The study of the degree of electronic entanglement that we have made in this paper is a useful resource for quantum computers whose input states are constructed in order to be maximally entangled. In this way, it is possible to obtain an exponential speed--up of quantum computation over classical computation.  
Therefore, the main aim of this work, is to analyze a dimer of ethylene, which represents the simplest organic conjugate system, in order to find the configuration corresponding to a maximum degree of entanglement between the two molecules. Thus we have to consider the compound system as a bipartite system where each molecule can be seen as a qubit and, consequently, calculate the degree of the entanglement due to the interaction of the molecules only, thus neglecting the internal interaction of each molecule.

In order to realize our goal we utilize the well known fact in quantum chemistry that the correlation energy between the two molecules of ethylene can be obtained as the difference between the correlation energy due to the interaction of all the electrons in the compound systems and the correlation energy of the electrons in each molecule, \emph{i.e.},
\begin{equation}
E_{c}^{int}=E_{c}[2C_{2}H_{4}]-2E_{c}[C_{2}H_{4}].
\end{equation}
Let us now observe that when the distance between the two $C_{2}H_{4}$ molecules is infinite, we can consider them as two separated subsystems of the compound system dimer, which we denote by $2C_{2}H_{4}$, since they are uncorrelated. In this case, we have that $\rho[2C_{2}H_{4}]=\rho[C_{2}H_{4}]\otimes\rho[C_{2}H_{4}]$, thus, by using the definition of von Neumann entropy, we have that $S(\rho[2C_{2}H_{4}])=2S(\rho[C_{2}H_{4}])$. In the general case, \emph{i. e.}, for finite disctances, the Klein's inequality holds, that is,
\begin{equation} 
S(\rho[2C_{2}H_{4}])>S(\rho[C_{2}H_{4}])+S(\rho[C_{2}H_{4}])=2S(\rho[C_{2}H_{4}]).
\end{equation}
 We thus define the \emph{interaction electron entanglement} ($S_{(int)}$) as
\begin{equation}
S_{(int)}=S(\rho[2C_{2}H_{4}])-2S(\rho[C_{2}H_{4}]).
\end{equation}
Then, we make a study of the degree of the interaction electron entanglement in the dimer of ethylene by changing the relative orientation and distance of the molecules. The   results obtained are reported in Figure 3, where it is shown the degree of interaction electron entanglement as a function of the rotation angle (in (a) and (b)) and as a function of molecular distance (in (c)) calculated in a face to face configuration. We can see that, for $R<3.1$ \AA,  $S_{int}$ grows up with the increasing of the molecular distance and the maximum value of this function moves from a rotation angle $\theta=50^{\circ}$ to $\theta=30^{\circ}$ and in general, the closer are the molecules the smaller is the rotation we would make to obtain the maximum value of $S_{int}$. Instead, for $R>3.1$ \AA, $S_{int}$ decreases with the increasing of $R$, and becomes less sensible to the rotation angle since the molecules are more and more uncorrelated. In order to confirm this behaviour, we plot $S_{int}$ as a function of $R$ for different torsion angles, in the Figure 3c. It is evident that the configuration obtained setting $R=3.1$ \AA $\,$ is the critical configuration. Indeed, for smaller distances, the entropy increases, while for bigger distances, it decreases with the increasing of $R$. Moreover, for $R=5.5$ \AA, the correlation between the two molecules is approximatively zero indipendently of the rotation angle.

\begin{tabular}{cp{5.5cm}}
\includegraphics[angle=-90, width=0.45\textwidth]{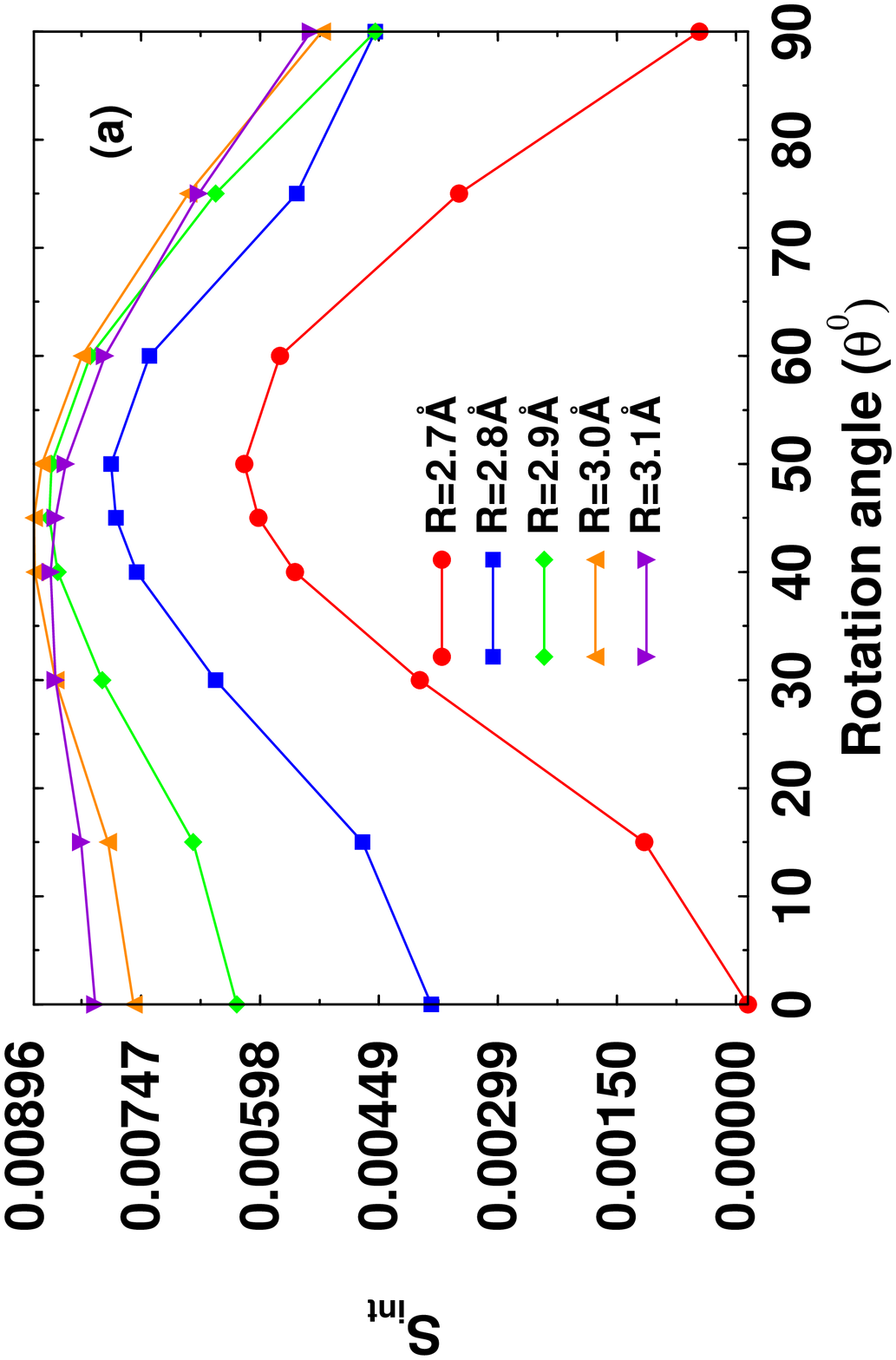}
&
\includegraphics[angle=-90, width=0.45\textwidth]{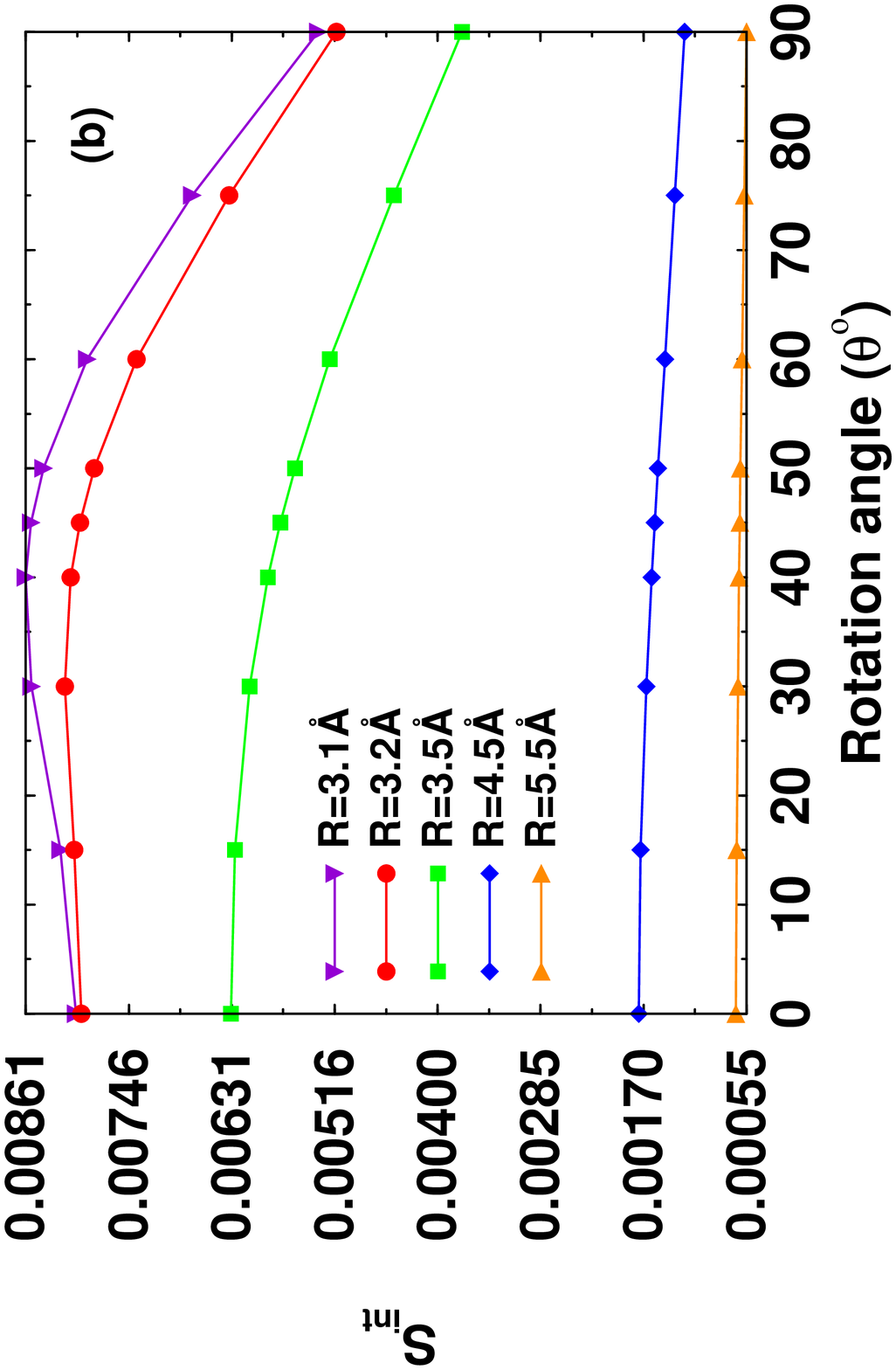}  \\
              &    \\
\includegraphics[angle=-90, width=0.45\textwidth]{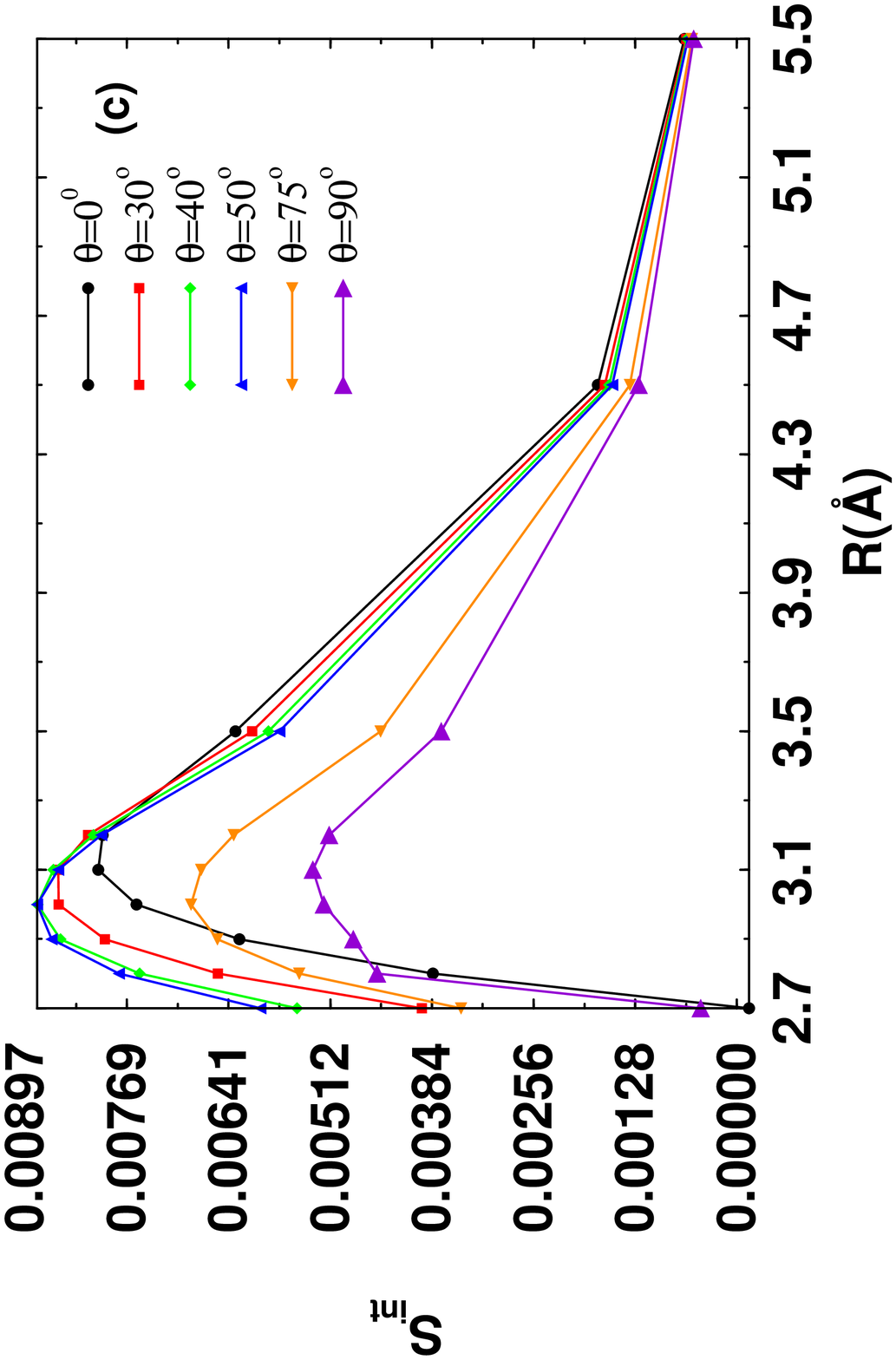}
&
Figure 3: {\it{Interaction electron entanglement for the dimer of ethylene as  function of rotation angle, calculated for different molecular planes distances (R) in a face to face configuration (in (a) and (b)). The same results are plotted as function of molecular distances for different rotation angles (in (c)).}}
\end{tabular}

\section{Computational details}
In order to calculate the {\small{CISD}} coefficients of the wave function, or equivalently, the entries of the reduced density matrix $\rho_{1}^{CISD}$ and to found the eigenvalues of the density matrix, the so called Natural Spin Orbitals, we use the package \emph {Gaussian 03} \cite{gau}. 
In the calculation concerning the dissociation of the $H_{2}$ molecule, we expand the wave function with a Configuration Interaction Single Double method. For this system, all the possible excitations are the single or the double ones, hence the CISD method represents a Full Configuration Interaction (FCI) method. In the study of the dimer of ethylene, we expand the wave function with a Coupled Cluster Single Double (CCSD) method \cite{ciz66}, \cite{scu88}. This is a numerical tecnique used for describing many--body systems and has the property of describing coupled two--body electron correlation effects. The CCSD method is an excitation truncated Coupled Cluster (CC) method constructed in an exponential approach by including only the desired excitation operators (single and double); hence it cannot be considered a FCI-like method. However it can describe the interaction of separated molecules better than CISD \cite{scu87}. For both systems, the input is prepared with an Unrestricted Hartree-Fock (UHF) calculation.
In fact, the UHF description of bond breaking in $H_{2}$ gives the proper dissociation products, while the Restricted Hartree-Fock (RHF) description of $H_{2}$ gives unrealistic ones. In the following we show that, in order to study the dissociation of $H_{2}$ molecule, the electron correlation energy must be defined by UHF appoximation as $|E_{corr}|=|E_{exact}-E_{UHF}|$. In fact, at short internuclear distances the RHF and UHF wave functions are identical but at large distances, outside the bond region, only UHF reproduces the correct progress of correlation energy of $H_{2}$ that must be zero when the two atoms are distant and each electron cannot interfere with the other (Figure 4).    
The correlation energy in Eq. (\ref{correlation}) is defined in terms of a complete one--electron basis. In practice, however, an incomplete basis must be used for the calculation of the correlation energy. The term \emph{correlation energy} is that used to denote the energy obtained from Eq. (\ref{correlation}) in a given one--electron basis.

The correlation energy usually increases in magnitude with the size of the orbital basis, since a small basis does not have the flexibility required for an accurate representation of correlation effects.

\begin{tabular}{cp{4.3cm}}           
\includegraphics[angle=-90, width=0.55\textwidth]{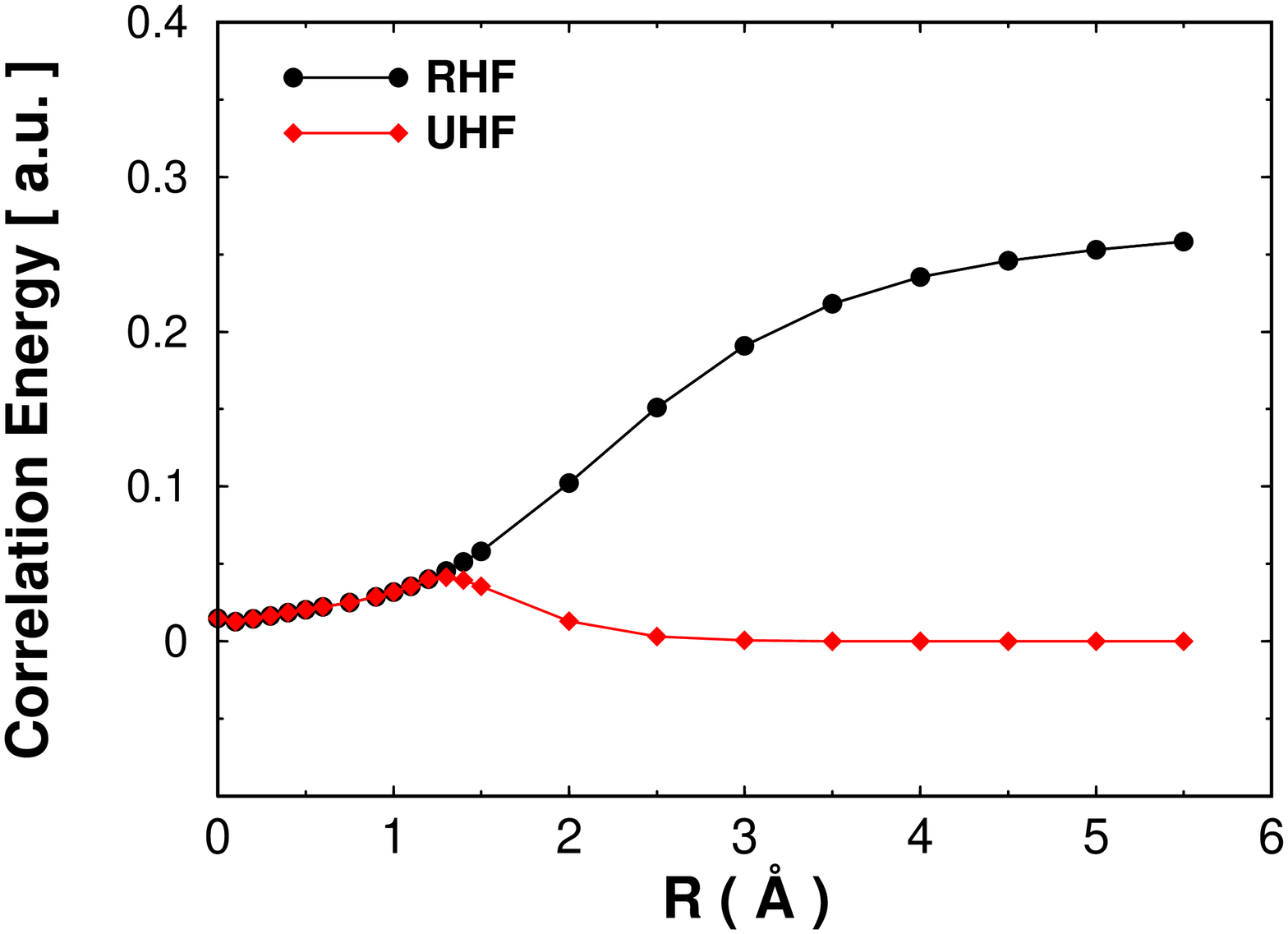}
&
{\it{Figure 4: The figure reports the electron correlation as a function of the internuclear distance R of the $H_{2}$ molecule using Gaussian basis set 3-21G with package Gaussian. The comparison between correlation energy calculated as $|E_{corr}|=|E_{exact}-E_{RHF}|$ and calculated as $|E_{corr}|=|E_{exact}-E_{UHF}|$ is showed. }}
\end{tabular}

In order to confirm this theory, we analyze four different kinds of basis sets known in the literature as $3-21G$, $6-31G$, $6-31G^{**}$, $6-31+$$+G^{**}$ and we note that the more increase the bases used the bigger becomes the contribute to correlation energy.

The basis sets $3-21G$ and $6-31G$ use $1s$, $2s$ and $1p$ atomic orbitals hence their contributions to the correlation energy are approximatively the same. The basis set $6-31G^{**}$ instead includes the orbital $1d$, while $6-31+$$+G^{**}$ uses the same orbitals but these are more diffuse. Hence, we can see that the energy correlation obtained with these two bases is bigger than the one obtained with the first basis sets. It is however important to note that at short internuclear distances the correlation energy strongly depends on the size of orbital basis whereas at large distances, outside the bond region, all four courses become approximatively the same.
  
After the above preliminary study of basis sets we decide to use the smallest basis set ($3-21G$). This choice allows us to reduce drastically the computational cost, which constitutes an enormus advantage in the study of other systems, in particular more complex molecules.
\section{Conclusions}
In this paper, we have analyzed and elaborated the suggestion in \cite{hua05} that the entanglement can be used as an alternative estimation of the electron correlation in quantum chemistry calculations.

We have, firstly, compared the degree of entanglement in the dissociation process of $H_{2}$ molecule calculated as the von Neumann entropy of the reduced density matrix $S(\rho_{1})$, as in \cite{hua05}, with the one calculated by the entropy of the density matrix of the compound system $S(\rho)$. The two behaviours obtained in this way are shown to be similar, hence, we have choosen to adopt $S(\rho)$ for a qualitative estimation of entanglement in more complex systems. Then, we have verified that the electron correlation energy for a dimer of ethylene is well reproduced by the entanglement $S(\rho)$ for different configurations of the system, as it has been shown in Figure 2.

Analyzing interacting molecules, we have introduced the interaction electron entanglement in order to calculate the only entanglement due to the interaction of the two molecules without their internal correlations. As it has been shown in Figure 3, there are configurations that maximize the degree of entanglement. This study of the degree of electronic entanglement is a useful resource for quantum computers whose input states are constructed in order to be maximally entangled. 

\section*{Appendix A}
The aim of this appendix is clarifying how one can calculate the reduced density matrix as that one we propose in Eq. (\ref{reddensH}). In order to semplify the calculations, we consider two electrons in a two--levels system.
Let us introduce the Hilbert space 
$\mathscr H=\Big[\mathscr L^{2}(1) \otimes \mathscr S^{2}(1)\Big]\otimes \Big[\mathscr L^{2}(2) \otimes \mathscr S^{2}(2)\Big]=\mathscr C^{4}(1)\otimes\mathscr C^{4}(2),$
where $\mathscr L$ and $\mathscr S$ represent the orbital and the spin space, respectively; in brackets we denote one of the two electrons and the apex represents the dimension of the space. An orthonormal basis for each space $\mathscr C^{4}$ in the occupation number representation $(n_{1}\uparrow, n_{1}\downarrow, n_{2}\uparrow, n_{2}\downarrow)$ is

$\left\{ \begin{array}{c}
|n_{1}\uparrow\rangle\\
|n_{1}\downarrow\rangle\\
|n_{2}\uparrow\rangle\\
|n_{2}\downarrow\rangle
\end{array}\right\}\otimes
\left\{ \begin{array}{c}
|n_{1}\uparrow\rangle\\
|n_{1}\downarrow\rangle\\
|n_{2}\uparrow\rangle\\
|n_{2}\downarrow\rangle
\end{array}\right\}$. 
A pure two-electron state $|\Psi\rangle$ in this case can be written as 
$|\Psi\rangle=\sum_{a=1}^{4}\sum_{b=1}^{4}\omega_{a,b}c_{a}^{\dag}c_{b}^{\dag}|0\rangle,$
where $|0\rangle$ is the vacuum state, the coefficient $\omega$, accordingly to Pauli exclusion principle, satisfies the following requests:
$\left\{ \begin{array}{l}
\omega_{a,b}=-\omega_{b,a}\\
\omega_{i,i}=0,
\end{array}\right.$
and $c^{\dag}$ and $c$ are the creation and annihilation operators of single--particle states, respectively whose action on the vacuum state is
\begin{displaymath}
c_{a}^{\dag}c_{b}^{\dag}|0\rangle = |ab\rangle,\hspace{0.3cm}
 a,b\in\{ 1,2,3,4\}\hspace{1.0cm}
\begin{tabular}{ll}
$1\equiv |n_{1}\uparrow \rangle$ & $3\equiv |n_{2}\uparrow \rangle$ \\
$2\equiv |n_{1}\downarrow \rangle$ & $4\equiv |n_{2}\downarrow \rangle$.\\
\end{tabular}
\end{displaymath}
Let us now consider the four sites instead of the two particles: they are single particle sites consequently they have no particle or one.
Since our aim is having some informations about the $n_{1}$ we proceed making the partial trace of the density operator $\rho$ with respect to $n_{2}\uparrow$ and $n_{2}\downarrow$ 
that is,

$\rho_{n_{1}}=Tr_{n_{2}}\rho=\sum_{\begin{tabular}{c}\begin{footnotesize} ${n_{2}\uparrow=0,1}$\end{footnotesize} \\ \begin{footnotesize}$n_{2}\downarrow=0,1$\end{footnotesize} \end{tabular}}\langle n_{2}\uparrow, n_{2}\downarrow|\rho|n_{2}\uparrow,n_{2}\downarrow\rangle=$

$=\langle00|\rho|00\rangle+\langle01|\rho|01\rangle+\langle10|\rho|10\rangle+\langle11|\rho|11\rangle$.

Let us calculate each matrix element separately.

The first one is
$\langle00|\rho|00\rangle=\sum_{a,b=1}^{4}\sum_{a',b'=1}^{4}\omega_{a,b}\omega^{*}_{a',b'}\langle 00|c_{a}^{\dag}c_{b}^{\dag}|0\rangle\langle0|c_{b'}c_{a'}|00\rangle$
where $a$, $b$, $a'$ and $b'$ assume the values $1,2,3,4$ and they represent the four sites of the two levels system that we are studying; moreover, since we make the inner product with $\langle 00|$ and $|00 \rangle$, the only elements of the summatory that have non--zero contribution are $|1100 \rangle$\footnote{In all these calculations, each vector with four components has the first and the second entries corresponding with each two sites of $n_{1}$ level and the third and fourth entries corresponding with $n_{2}$ level. Hence the ket $|1100\rangle$, for example, represents two electrons on the level $n_{1}$ and no one on the level $n_{2}$.} thus they correspond to $\omega_{1,2}$ and $\omega_{2,1}$:

\begin{tabular}{ll}
$|1100 \rangle$:& $|n_{1}\uparrow\rangle\otimes|n_{1}\downarrow\rangle\rightarrow \omega_{12}$\\
& $|n_{1}\downarrow\rangle\otimes|n_{1}\uparrow\rangle\rightarrow \omega_{21}.$
\end{tabular}

Hence
$\langle00|\rho|00\rangle=\sum_{a,b=1}^{2}\sum_{a',b'=1}^{2}\omega_{a,b}\omega^{*}_{a',b'}\langle 00|c_{a}^{\dag}c_{b}^{\dag}|0\rangle\langle0|c_{b'}c_{a'}|00\rangle$.
Making explicit the sum and using the fact that $\omega_{ii}=0$, it becomes:
\begin{displaymath}
\langle00|\rho|00\rangle=\omega_{1,2}\omega_{1,2}^{*}\langle00|c_{1}^{\dag}c_{2}^{\dag}|0\rangle \langle 0|c_{2}c_{1}|00\rangle+\omega_{2,1}\omega_{1,2}^{*}\langle00|c_{2}^{\dag}c_{1}^{\dag}|0\rangle \langle 0|c_{2}c_{1}|00\rangle+
\end{displaymath}
\begin{equation}
\label{sommaesplicita}
+\omega_{1,2}\omega_{2,1}^{*}\langle00|c_{1}^{\dag}c_{2}^{\dag}|0\rangle \langle 0|c_{1}c_{2}|00\rangle+\omega_{2,1}\omega_{2,1}^{*}\langle00|c_{2}^{\dag}c_{1}^{\dag}|0\rangle \langle 0|c_{1}c_{2}|00\rangle.
\end{equation}
The action of the creation and distruction operators on the vacuum state produces some states such as $\pm|n_{1}\uparrow n_{1}\downarrow\rangle$ and $\pm|n_{1}\downarrow n_{1}\uparrow\rangle$ where the double sign depends on the order of the operators; using no more the representation refering to fermions but the one refering to the sites all these states can be represented with $\pm|1100\rangle$ so Eq (\ref{sommaesplicita}) becomes:
 \begin{displaymath}
\langle00|\rho|00\rangle=\omega_{1,2}\omega_{1,2}^{*}\langle00|1100\rangle \langle 1100|00\rangle-\omega_{2,1}\omega_{1,2}^{*}\langle00|1100\rangle \langle 1100|00\rangle+
\end{displaymath}
\begin{equation}
-\omega_{1,2}\omega_{2,1}^{*}\langle00|1100\rangle \langle 1100|00\rangle+\omega_{2,1}\omega_{2,1}^{*}\langle00|1100\rangle \langle 1100|00\rangle.
\end{equation}
Using the orthogonality, we obtain:
$\langle00|\rho|00\rangle=4|\omega_{12}|^{2}|11\rangle\langle11|$.
Let us analyze the second matrix element:

$\langle01|\rho|01\rangle$
$=\sum_{a,b=1}^{4}\sum_{a',b'=1}^{4}\omega_{a,b}\omega^{*}_{a',b'}\langle 01|c_{a}^{\dag}c_{b}^{\dag}|0\rangle\langle0|c_{b'}c_{a'}|01\rangle.$
Since we make the inner product with $\langle 01|$ and $|01 \rangle$, the only elements of the summatory that have non-zero contribution are $|1100 \rangle$ thus they correspond to $\omega_{1,2}$ and $\omega_{2,1}$:
\begin{tabular}{ll}
$|1001 \rangle$: & $|n_{1}\uparrow\rangle\otimes|n_{2}\downarrow\rangle\rightarrow \omega_{14}$\\
& $|n_{2}\downarrow\rangle\otimes|n_{1}\uparrow\rangle\rightarrow \omega_{41}$\\
$|1001 \rangle$: & $|n_{1}\downarrow\rangle\otimes|n_{2}\downarrow\rangle\rightarrow \omega_{24}$\\
& $|n_{2}\downarrow\rangle\otimes|n_{1}\downarrow\rangle\rightarrow \omega_{42}.$
\end{tabular}

Hence
$\langle01|\rho|01\rangle=\sum_{a,b=1,2,4}\sum_{a',b'=1,2,4}\omega_{a,b}\omega^{*}_{a',b'}\langle 01|c_{a}^{\dag}c_{b}^{\dag}|0\rangle\langle0|c_{b'}c_{a'}|01\rangle$. 
\\
Making explicit the sum above and using the site representation as before, this matrix element becomes:
$\langle01|\rho|01\rangle=4|\omega_{1,4}|^{2}|10\rangle\langle10|.$
Now, for the third matrix element, that is:
$\langle10|\rho|10\rangle=\sum_{a,b=1}^{4}\sum_{a',b'=1}^{4}\omega_{a,b}\omega^{*}_{a',b'}\langle 10|c_{a}^{\dag}c_{b}^{\dag}|0\rangle\langle0|c_{b'}c_{a'}|10\rangle$
we note that the only vectors that have a non zero contribution in the summartory after the inner product with $\langle 10|$ and $|10 \rangle$ are:

\begin{tabular}{ll}
$|1010 \rangle$: & $|n_{1}\uparrow\rangle\otimes|n_{2}\uparrow\rangle\rightarrow \omega_{13}$\\
& $|n_{2}\uparrow\rangle\otimes|n_{1}\uparrow\rangle\rightarrow \omega_{31}$\\
$|0110 \rangle$: & $|n_{1}\downarrow\rangle\otimes|n_{2}\uparrow\rangle\rightarrow \omega_{23}$\\
& $|n_{2}\uparrow\rangle\otimes|n_{1}\downarrow\rangle\rightarrow \omega_{32}.$
\end{tabular}

Hence, $\langle10|\rho|10\rangle=4|\omega_{2,3}|^{2}|10\rangle\langle10|$. Finaly, the latest matrix element, is, analogously,  $\langle11|\rho|11\rangle=4|\omega_{3,4}|^{2}|00\rangle\langle00|$

Thus, in the basis $\Big \{ |00\rangle;|01\rangle;|10\rangle; |11\rangle  \Big \}$, $\rho_{n_{1}}$ reads
\begin{equation}
\label{densitan1}
\rho_{n_{1}}=\left( \begin{array}{cccc}
4|\omega_{3,4}|^{2} & 0 & 0 & 0\\
0 & 4|\omega_{2,3}|^{2} & 0 & 0\\
0 & 0 & 4|\omega_{1,4}|^{2} & 0\\
0 & 0 & 0 & 4|\omega_{1,2}|^{2}
\end{array} \right).
\end{equation}
Equation (\ref{densitan1}) represents the density operator for the level $n_{1}$ and it is important to spend some words on the entryes of the matrix: $\omega_{3,4}$ represents the excitations on the second level so, in the two level system that we consider, $\omega_{3,4}$ is associated with the probability that $n_{1}$ is empty;  $\omega_{1,2}$ is associated with the possibility that the two electrons are in $n_{1}$, the coefficients $\omega_{2,3}$ and $\omega_{1,4}$ are associated with the probability that $n_{1}$ can be occupated by one eletron. This fact allows us to introduce the following notation:
$\rho_{n_{1},0}\equiv4|\omega_{3,4}|^{2};\hspace{0.5cm}
\rho_{n_{1},2}\equiv4|\omega_{1,2}|^{2};\hspace{0.5cm}
\rho_{n_{1},1}\equiv\left(\begin{array}{cc}
4|\omega_{2,3}|^{2} & 0\\
0 & 4|\omega_{1,4}|^{2} \\
\end{array} \right),$
where $\rho_{n_{1},0}$ denoted an empty orbital, $\rho_{n_{1},2}$ denotes  two accupied orbitals and $\rho_{n_{1},1}$ denotes one electron occupied orbital.
In order to obtain some informations about one of the two electrons, we have to make tha partial trace of $\rho_{n_{1}}$ respect one of the sites of the level $n_{1}$ ($n_{1}\uparrow, n_{1}\downarrow$):

$\rho_{1}=Tr_{n_{1}\uparrow}\rho_{n_{1}}=\sum_{n_{1}\uparrow=0,1}\rho_{n_{1}}
=\left( \begin{array}{cc}
4 \Big(|\omega_{3,4}|^{2}+|\omega_{2,3}|^{2}\Big) & 0 \\
0 & 4 \Big(|\omega_{1,4}|^{2}+|\omega_{1,2}|^{2} \Big)  
\end{array} \right).$ 
In terms of CISD expansion coefficients, the transition amplitude $\omega_{i,j}$ can be written as $|\omega_{1,2}|^{2}=\frac{c_{0}^{2}}{4}$, $|\omega_{1,4}|^{2}=\frac{c_{2}^{2}}{4}$, $|\omega_{3,4}|^{2}=\frac{c_{1,2}^{2}}{4}$ and $|\omega_{2,3}|^{2}=\frac{c_{1}^{2}}{4}$, hence $\rho_{1}$ becames:
$\rho_{1}^{CISD}=\left( \begin{array}{cc}
|c_{1}|^{2}+|c_{1,2}|^{2} & 0\\
0 & |c_{0}|^{2}+|c_{2}|^{2}
\end{array} \right).$    
Eq (\ref{reddensH}) is obtained as a generalization of this model to a \emph{m--}level system.


\begin{thebibliography}{99}
\bibitem{nie00} Nielsen and Chuang \emph{Quantum computation and quantum information}, Cambridge Univ. Press, Cambridge, 2000.
\bibitem{ben93} C. H. Bennet, \emph{et al.}, Phys. Rev. Lett. \textbf{70} (1993) 1895.
\bibitem{ben92} C. H. Bennet and S. J. Wiesner, Phys. Rev. Lett. \textbf{69} (1992) 2881.
\bibitem{eke91} A. K. Ekert, Phys. Rev. Lett \textbf{67} (1991) 661.
\bibitem{fuc97} C. A. Fuchs, Phys. Rev. Lett. \textbf{79} (1997) 1162.
\bibitem{gho03} S. Ghosh, \emph{et al.}, Nature \textbf{425}, (2003) 48.
\bibitem{che06} Y. Chen, \emph{et al.}, quantum--ph/ 0407228 (2006).
\bibitem{lix05} L. He, G. Bester, and A. Zunger, cond--math/0503492 (2005).
\bibitem{bus06} F. Buscemi, P. Bordone, and A. Bertoni quantum--ph/ 0602127 (2006).
\bibitem{sch01} J. Schliemann \emph{et al.}  Phys. Rev .\textbf{A 64}, (2001) 022303. 
\bibitem{ghi04} G. C. Ghirardi and L. Marinatto, Phys. Rev. \textbf{A 70}, (2004) 012109.
\bibitem{hua05} Z. Huang, S. Kais Chem. Phys. Lett. \textbf{413} (2005) 1-5.
\bibitem{nat93} Z. Naturforsch Phys. Rev. \textbf{A 56}, (1997) 4477.
\bibitem{ram97} J. C. Ramirez. et al. Phys. Rev. \textbf{A 56} (1997) 447. 
\bibitem{sza89} A. Szabo, N. S. Ostlund \emph{Modern Quantum Chemistry: Introduction to Advanced Electronic Structure Theory} McGraw Hill (1989).
\bibitem{sch35} E. Schr\"odinger, Naturwissenschaften \textbf{23} (1935) 807.
\bibitem {doh02} A. C. Doherty, \emph{et al.}, Phys. Rev. Lett. \textbf{88} (2002) 187904.
\bibitem{des76} B. D'Espagnat, \emph{Conceptual Foundations of Quantum Mechanics} (Benjamin, Reading, MA, 1976).
\bibitem{von32} J. von Neumann, \emph{Mathematical Foundations of Quantum Mechanics}, (Prinveton University Press, Princeton, 1955).
\bibitem{git02} J. R. Gittings, and A. J. Fisher Phys. Rev. \textbf{A 66}, (2002) 032305.
\bibitem{pal82} J. Paldus, \emph{et al.}, J. Chem. Phys. \textbf{76}, (1982) 2458--2470.
\bibitem{scu87} G. E. Scuseria, A. C. Scheiner, T. J. Lee, J. E. Rice, H. F. Schaefer J. Chem. Phys. \textbf{86} (1987) 2881.
\bibitem{zan02} P. Zanardi, Phys. Rev. \textbf{A 65} (2002) 042101.
\bibitem{ben96} C. H. Bennett, H. Bernstein, S. Popescu, and B. Schumacher, Phys. Rev \textbf{A 53}, (1996) 2046.
\bibitem{got05} A. D. Gottlieb and N. J. Mauser, Phys. Rev. Lett. \textbf{95} (2005) 123003.
\bibitem{gro94} R. Grobe, K. Rzazewski, and J. H. Eberly, J. Phys. \textbf{B 27},(1994) L503.
\bibitem{ger97} P. Gersdorf, \emph{et al.}, Int. J. Quantum Chem. \textbf{61}, (1997) 935. 
\bibitem{gau} M. J. Frish et al., GAUSSIAN 98, Revision A. 11.3, Gaussian, Inc., Pittsburg PA, 1998.
\bibitem{ciz66} J. Cizek, J. Chem. Phys. \textbf{45}, (1966) 4256.
\bibitem{scu88} G. E. Scuseria, \emph{et al.}, J. Chem. Phys. \textbf{89}, (1988) 7382.
\bibitem{fel93} D. Feller, E. D. Glendening et al., J. Chem. Phys. \textbf{99}, (1993) 2829.



\end{thebibliography}
\end{document}